\documentclass[a4paper,12pt]{article}
\usepackage{amssymb,amsmath}
\usepackage{epsfig,fancyhdr}
\usepackage{a4}
\usepackage{parskip}
\usepackage{appendix}
\usepackage{color,soul}
\usepackage{amsthm,amsxtra,amscd,relsize,multirow}
\usepackage[all,2cell]{xy}\UseAllTwocells

\newcommand{\sect}[1]{\setcounter{equation}{0}\section{#1}}
\renewcommand{\theequation}{\arabic{section}.\arabic{equation}}

\def\N{{\mathcal N}}
\def\L{{\mathcal L}}
\def\E{{\mathcal E}}

\def\J{{\mathcal J}}

\def\L{{\mathcal L}}

\def\ds{\displaystyle}

\def\a{\alpha}
\def\b{\beta}

\def\g{\gamma}

\def\s{\sigma}
\def\t{\tilde}

\def\m{\mu}

\def\f{\phi}
\def\vf{\varphi}

\def\l{\lambda}

\def\o{\omega}

\def\cosh{\mathrm{cosh}}

\def\p{\partial}
\def\rb{\right}
\def\lb{\left}
\def\axs{AdS_5\times S^5}
\newcommand{\eq}[1]{\begin{equation} #1 \end{equation}}
\newcommand{\al}[1]{\begin{align} #1 \end{align}}
\newcommand{\ml}[1]{\begin{multline} #1 \end{multline}}

\begin{titlepage}
\title{Strings on the deformed $T^{1,1}$: giant magnon and single spike solutions}

\author{
H. Dimov${}^{\dagger}$, M. Michalcik${}^{\ddagger}$ and 
R.C. Rashkov${}^{\dagger,\ddagger}$\thanks{e-mail:
rash@hep.itp.tuwien.ac.at.}
\ \\ \ \\
${}^{\ddagger}$ Institute for Theoretical Physics, \\ Vienna
University of Technology,\\
Wiedner Hauptstr. 8-10, 1040 Vienna, Austria 
\ \\ \ \\
${}^{\dagger}$  Department of Physics, Sofia
University,\\
5 J. Bourchier Blvd, 1164 Sofia, Bulgaria
}
 \end{titlepage}

\begin{document}

\maketitle
\thispagestyle{fancy}

 \begin{abstract}
In this paper we find giant magnon and single spike string solutions in a sector of the gamma-deformed conifold. We examine the dispersion relations and find a behavior analogous to the undeformed case. 
The transcendental functional relations between the conserved charges are shifted by certain gamma-dependent term. The latter is proportional to the total momentum and thus qualitatively different from known cases. 
 \end{abstract}

\sect{Introduction}

One of the most exciting topics in the high energy theory over the last decades is the correspondence between strings and gauge fields. One of the most promising explicit realizations of this correspondence 
was provided by the Maldacena conjecture about AdS/CFT correspondence \cite{holography}. 

The semi-classical string has played an important role in studying various aspects of the
$AdS_5/SYM_4$ correspondence \cite{Bena:2003wd}-\cite{Lee:2008sk}.
The developments and successes in this
particular case suggest the methods and tools that should be used to investigate the new emergent duality. 
The best studied example of the duality between strings and gauge theory is
the AdS/CFT correspondence on $AdS_5 \times S^5$.
One of the most important predictions of the correspondence is the equivalence
between the spectrum of string theory and the spectrum of anomalous dimensions of gauge invariant
operators. There has been a good deal of success recently
in comparing the energies of semiclassical strings and the
anomalous dimensions of the gauge theory operators.

An important, even conceptual role in these studies played the integrability. 
Assuming that these theories are integrable, the dynamics
should be encoded in an appropriate scattering matrix
S. This can be interpreted from both sides of the correspondence
as follows. On the string side, in the strong coupling
limit the S matrix can be interpreted as describing
the two-body scattering of elementary excitations on the
world sheet. When their world-sheet momenta become
large, these excitations can be described as special types
of solitonic solutions, or giant magnons, and the interpolating
region is described by the dynamics of the so-called
near-flat-space regime \cite{Hofman:2006xt,MS,BT}. On the gauge theory side, the
action of the dilatation operator on single-trace gauge-invariant operators is
the same as that of a Hamiltonian acting on the states of a
certain spin chain \cite{Minahan:2002ve}. This turns out to be of great advantage
because one can diagonalize the matrix of anomalous
dimensions by using the ``magic'' algebraic Bethe ansatz technique.
 Insertion of different operators into the single trace long operators is interpreted 
as magnons and the S-matrix factorizes to two-magnon scatterings governing the spectrum. 

On the string theory side, the corresponding classical string solutions are called giant magnons 
and have the shape of arcs moving along some isometry direction. The angle deficit defined by 
the end pints of the arcs is identified as the momentum of the magnon, while the dispersion 
relations determine the anomalous dimension of a certain gauge theory operator at strong coupling. 

Another important string solution is the so called single spike string. Its shape has only one 
single spike and large winding number in some isometry direction. 
While the giant magnon solutions can be interpreted as higher twist operators
in the field theory, the single spike solutions do not
seem to be directly related to some field theory operators.
However, in \cite{Ishizeki:2007we} an interpretation of this solution as a spin
chain Hubbard model, which means the antiferromagnetic
phase of the corresponding spin chain, was found, but the
relation to field theory operators is still unclear.
Although not completely understood, the spiky string solutions is believed to play important 
role in the AdS/CFT correspondence.

After the impressive achievements in the most supersymmetric example of AdS/CFT correspondence,
namely $AdS_5 \times S^5$, it is important to extend the considerations to less supersymmetric 
gauge theories, more over that the later are more interesting from physical point of view. The integrability is expected to play the same crucial role, but unfortunately not much is known on 
the subject. There are several ways to find a theory with less supersymmetry. One of them is 
suggested by Lunin and Maldacena \cite{Lunin:2005jy}. The authors consider $\N=4$  supersymmetric Yang-Mills (SYM) and its marginal deformations \cite{Leigh:1995ep}. In \cite{Lunin:2005jy} the supergravity dual of marginaly deformed supersymmetric Yang-Mills theory has been identified. 
Latter an explicit deforming procedure,called TsT transformation, and the integrability of the resulting backgrounds has been presented in \cite{Frolov:2005dj}.
Giant magnons and single spike strings has been studied in \cite{Chu:2006ae}-\cite{Bykov:2008bj}.
In \cite{Chu:2006ae,Bobev:2006fg} the deformation parameter enters the dispersion relation for the giant magnon as a shift by $\pi\gamma$. In \cite{Bykov:2008bj} it was argued that, in the limit of the conserved charge  $\J=J/g\rightarrow\infty$ and upon the  identification $\gamma\sqrt{\l}\sim \hat\gamma$, the shift by $\gamma$ should not be seen by the classical theory. It is important however that the deformation introduces a non-trivial twist in the boundary conditions for the isometry directions with essential consequences.

There is another way to approach less supersymmetric backgrounds. The experience from the AdS/CFT correspondence suggests that one can take a stack of N D3 branes and place them not in a flat 
space, but at the apex of a conifold \cite{Klebanov:1998hh}. This model possess a lot of 
interesting 
features and allows to build gauge theory operators of great physical importance. The resulting 
ten dimensional space time takes form of the direct product $AdS_5\times T^{1,1}$. Since then an 
infinite families of five dimensional spaces, called Sasaki-Einstein spaces, complementing $AdS_5$ space have been constructed \cite{Gauntlett:2004yd}-\cite{Cvetic:2005ft} as well as their gauge 
theory duals were identified \cite{Benvenuti:2004dy}-\cite{Butti:2005sw}. Further developments 
can be traced in \cite{Gauntlett:2004yd}-\cite{Benvenuti:2008bd}.

The powerful solution generating technique based on Lunin-Maldacena construction has been applied 
to various backgrounds in \cite{CatalOzer:2005mr}. 
The deformations in this paper also include conifold which is certainly of interest for AdS/CFT correspondence.

Inspired by the considerations in \cite{Benvenuti:2008bd} and \cite{Kim:2003vn,Dimov:}, we investigate 
the giant magnon and the single spike string solutions in beta-deformed conifold background. The dispersion relations are supposed to describe the anomalous dimensions of particular class of 
gauge theory operators. We expect our results to shed some light on the gauge theory description 
of the  conjectured duality as well as integrability of certain subsectors of the theory.

The paper is organized as follows. In the second Section we review the result of \cite{Benvenuti:2008bd}, beta-deformations and giant magnons and spiky strings in such backgrounds.
 Section 3 presents the string theory in a consistently truncated subsector of $T^{1,1}$. 
In Section 4 we derive the dispersion relations for the cases of giant magnon and single spike 
string solutions. The obtained results are summarized in Conclusions. Some helpful formulae are presented in an Appendix.


\sect{Review of the known results}

In this Section we review first the giant magnon  and single spike strings on the conifold and 
then briefly describe the same issues in the beta-deformed $S^5_\gamma$ (here $\gamma=\Re\beta$).
We fix here some notations and present methods which  will used in what follows.

\subsection{Giant magnons and single spike strings on the conifold}

Here we will briefly review the results of \cite{Benvenuti:2008bd}.
The metric of the conifold can be written as $dr^2+r^2\:d\Omega^2_{T^{1,1}}$ and 
combined with the metric of a stack of N D3 branes can be organized as $AdS_5\times T^{1,1}$.
The description best suited for our purpose is as follows.
Let us consider strings moving in $T^{1,1}$, which is
a homogeneous space $(SU(2)\times SU(2))/U(1)$, with $U(1)$ chosen
to be a diagonal subgroup of the maximal torus in $SU(2)\times
SU(2)$. One can start with more general set up of squashed spheres employed in \cite{Benvenuti:2008bd} 
with an  explicit form of the metric written as 
$U(1)$ bundle over $S^2\times S^2$, 
\al{ ds^2 &= a (d\theta_1^2 +
\sin^2 \theta_1 d\phi_1^2 +
  d\theta_2^2 + \sin^2 \theta_2 d\phi_2^2)
\nonumber\\
&+ b( d\psi + p\cos\theta_1 d\phi_1 + q\cos\theta_2 d\phi_2)^2 .
 \label{metric} 
}
Here $\theta_i,\phi_i$ are the coordinates
of the two $S^2$, and the $U(1)$ fiber is parameterized by $\psi \in [0,4\pi]$.
The space is an Einstein manifold if the following choice of the parameters is made
$a=\frac{1}{6},b=\frac{1}{9}$. Supersymmetry requirements further 
restricts $p=q=1$ and the
the space becomes supersymmetric, i.e. the resulting Sasaki-Einstein manifold allows two Killing spinors, hence $\N=1$ supersymmetry.

The part $T^{1,1}$ provides the angular
part of a singular Calabi-Yau manifold. One can easily see from
Eq.~(\ref{metric}) that the isometry is $SU(2)\times SU(2)\times
U(1)$. The three mutually commuting Killing vectors can be chosen
as $\partial_{\phi_1},\partial_{\phi_2}, \partial_\psi$.

We will proceed however with the choice $p=q=1$ but with squashing parameter $b$ unfixed 
($a=b/4$). It was shown in \cite{Benvenuti:2008bd} that one can consistently set, say $\theta_2,\f_2=const$.
The starting point then is the subspace of $T^{1,1}$ corresponding to the metric
\eq{
ds^2=-dt^2+\frac{b}{4}\Big[d\theta^2+\sin^2\theta\:d\f^2+
b\big(d\psi-\cos\theta\:d\f\big)^2\Big],
}
where the time coordinate $t\in \mathbb{R}$ originates from $AdS_5$.

\paragraph{Equations}\

Let us consider the sector defined by $\theta_2,\f_2=const$. To obtain solitonic solutions we 
use the ansatz
\eq{
t=\kappa\tau, \:\theta\equiv\theta_1=\theta(y),\:
\Psi=\o_\psi\tau+\psi(y), \:\Phi=\o_\f\tau+\f(y),
}
where $y=-d\tau+c\sigma$, $\Psi$ describes the $U(1)$ fiber and $\Phi\equiv\f_1$.

Integrating once the equations for the angles $\Psi$ and $\Phi$ in terms of $\theta$ 
and using the Virasoro constraints one finds
\eq{
u'=4\big[a_4u^4+a_3u^3+a_2u^2+a_1u+a_0\big].
\label{gen-eq-1}
}
where $u=\cos^2\theta/2$. Imposing appropriate boundary conditions we end up with
($\a_\pm >0$).
\eq{
u'^2=\a^2\o_\f u^2\big(\a_+-u\big)\big(u+\a_-\big).
\label{gen-eq-2}
}
The following relations between the integration constants and frequencies determine the
profile of the solution ($A_\psi=A_\f$)
\al{
& A_\f=\frac{d}{9}\big(\o_\psi+\o_\f\big)\quad \text{giant magnon}\\
& A_\f=\frac{c^2}{9d^2}\big(\o_\psi+\o_\f\big)\quad \text{single spike}.
}

\paragraph{Dispersion relations}\

For magnon type and spiky string solutions the conserved charges are
\al{
& P_t=-T\frac{b}{2}(\o_\psi+\o_\f) \\
& P_\psi=Tb\Big(\frac{b}{4}(\o_\psi+\o_\f)-\frac{b\o_\f}{2\big(1-d/c)^2\big)}u(y)\Big)\\
& P_\f=Tb\Big(\frac{b}{4}(\o_\psi+\o_\f)-\frac{b\o_\f}{2\big(1-(d/c)^2\big)}
\Big(2(1-b)u(y)^2+\big(b\Omega-2(1-b)\big)u(y)\Big),
}
where $u(y)=\cos^2\theta/2$ and $\Omega=(1-b)/b$.

The finite quantities giving  the dispersion relations are
\eq{
\E-\frac{2}{b}\J_\psi, \quad \E-\frac{2}{b}\J_\f,\quad
\E-\frac{\J_\psi+\J_\f}{b}, \quad
\quad \frac{\J_\psi-\J_\f}{b}.
}

The \textbf{giant magnon} dispersion relations on the conifold are
\eq{
\frac{\sqrt{3}}{2}\big(\E-3\J_\psi\big)=
\frac{\sqrt{3}\big(\E-3\J_\psi\big)/2-\cos\Delta\f}
{\sin\big(\sqrt{3}(\E-3\J_\psi)/2\big)}.
\label{disp-magn11}
}

Note that the dispersion relations are quite different from those in 
the most supersymmetric case.

The \textbf{single spike} string solutions obey the following dispersion
relations
\eq{
\frac{3\sqrt{3}J_\psi}{2}=\frac{\cos(3\sqrt{3}J_\psi)-\cos(2/3E-\Delta\vf)}
{\sin(3\sqrt{3}J_\psi)}.
\label{disp-spike11}
}

Again the dispersion relations are quite different from those in 
the most supersymmetric case, namely they have transcendental functional dependence 
between the charges.

\paragraph{Gauge theory side}\

The dual conformal field theory is known as the Klebanov-Witten model \cite{Klebanov:1998hh} and
is constructed considering a stack of D3 branes which are placed at the tip of a conifold.

The dual conformal field theory is identified as $\N=1$ supersymmetric $U(N)\times U(N)$
gauge theory with two chiral multiplets $A_i$ in $(N,\overline{N})$ and
another two, usually denoted by $B_i$, in $(\overline{N},N)$.
The angular part of the conifold is $T^{1,1}$ and its isometries determine the global symmetries of the gauge theory.
Being $U(1)$ bundle over $S^2\times S^2$,
this theory obviously has $SU(2)\times SU(2)$ global symmetry
which act separately on the doublets $A_i,B_i$,
and also a non-anomalous $U(1)$ R-symmetry.

The most general superpotential which respect the $SU(2)\times SU(2)\times U(1)_R$ symmetry 
 is a quartic superpotential of the form
\eq{
W =  \frac{g}{2} \epsilon^{ij} \epsilon^{kl}  \mathrm{Tr} A_i B_k A_j B_l .
}
Note that there is also $\mathbf{Z}_2$ symmetry. In the geometric picture, i.e. 
on conifold, it acts as reflection and on the gauge theory point of view it exchanges 
the two pairs $A_i$ and $B_j$.

The AdS/CFT correspondence suggests that the anomalous dimension of the gauge theory 
operators are encoded in the dispersion relation in the string theory. Therefore, 
here we are interested primarily in the conserved quantities which are
the energy $E  =  \sqrt{\lambda} \kappa$  and the following three angular momenta,
\eq{
J_A  \equiv   P_{\phi_1},
\quad
J_B \equiv  P_{\phi_2},
\quad
J_R \equiv   P_\psi. 
}

In order to have reliable comparison we must consider long composite operators constructed out of
$A_i$ and $B_j$. Then, it is natural to suggest
a correspondence between quantum numbers in string theory and
the dual operators.
 As it was shown in \cite{Klebanov:1998hh}, strings moving in $T^{1,1}$
are dual to pure scalar operators, i.e. they do not
contain fermions, covariant derivatives or gauge field strengths. One can construct scalar by making use of the fact that they are in the bi-fundamental representation. Therefore, the gauge singlets have the form
\eq{
\mathrm{Tr}\Big(A\:B\cdots A\:\bar A \cdots \bar B\:B\cdots \bar B\:\bar A \cdots\Big).
}
This form of the operators suggests the correspondence
\al{
J_A & \longleftrightarrow \frac{1}{2}
\Big[
\#(A_1)- \#(A_2) + \#(\overline{A}_2) -\#(\overline{A}_1)
\Big]
\\
J_B & \longleftrightarrow \frac{1}{2}
\Big[
\#(B_1)- \#(B_2) + \#(\overline{B}_2) -\#(\overline{B}_1)
\Big]
\\
J_R & \longleftrightarrow \frac{1}{4}
\Big[
\#(A_i)+\#(B_i) - \#(\overline{A}_i)-\#(\overline{B}_i)
\Big]
}
where  $\#(A_1)$ is the number of $A_1$'s under the trace of the dual composite operator etc.

We note that there exists an inequality between the
bare dimension and the $R$-charge, which is quite natural when 
written in terms of the string variables,
\eq{
E \geq 3| J_R |
\, .
\label{bound}
}
On gauge theory side it comes from the unitarity bound of $\N=1$ superconformal algebra. When the bound is saturated the primary fields close a chiral ring.
Complete dictionary between conserved charges in string theory and the dual gauge theory operator remains an open problem.

The derivation of the general string solution is a subject to much more complicated task 
related to issues as integrability etc.


\subsection{Giant magnons and single spike strings on $S^3_\gamma$}

Here we review the $\b$-deformed $\axs$ background found by Lunin
and Maldacena \cite{Lunin:2005jy}. This background is conjectured
to be dual to the Leigh-Strassler marginal deformations of $\N=4$
SYM \cite{Leigh:1995ep}. We note that this background can be
obtained from pure $\axs$ by a series of TsT transformations as
described in \cite{Frolov:2005dj}. The deformation parameter
$\b=\g+i\sigma_d$ is in general a complex number, but in our
analysis we will consider $\s_d=0$, in this case the deformation
is called $\gamma$-deformation. The resulting supergravity
background dual to real $\b$-deformations of $\N=4$ SYM is:
\eq{
ds^2=R^2\left(ds^2_{AdS_5}+\ds\sum_{i=1}^{3}(d\m_i^2+G\m_i^2d\phi_i^2)+
\t{\g}^2G\m_1^2\m_2^2\m_3^2(\ds\sum_{i=1}^3d\phi_i^2)\right)
\label{3.1}}
This background includes also a dilaton field as well as RR and
NS-NS form fields. The relevant form for our classical string
analysis will be the antisymmetric B-field:
\eq{
B=R^2\t{\g}G\left(\m_1^2\m_2^2d\phi_1d\phi_2+\m_2^2\m_3^2d\phi_2d
\phi_3+\m_1^2\m_3^2d\phi_1d\phi_3\right)
\label{3.2}}
In the above formulae we have defined
\eq{\begin{array}{l} \t{\g}=R^2\g \qquad\qquad R^2=\sqrt{4\pi
g_sN}=\sqrt{\l}
\\\\
G=\ds\frac{1}{1+\t{\g}^2(\m_1^2\m_2^2+\m_2^2\m_3^2+\m_1^2\m_3^2)}\\\\
\m_1=\sin\theta\cos\psi \qquad \m_2=\cos\theta \qquad
\m_3=\sin\theta\sin\psi
\end{array}\label{3.3}}
Where ($\theta$,$\psi$,$\phi_1$,$\phi_2$,$\phi_3$) are the usual
$S^5$ variables. This is a deformation of the $AdS_5\times S^5$
background governed by a single real deformation parameter
$\tilde{\gamma}$ and thus provides a useful setting for the
extension of the classical strings/spin chain/gauge theory duality
to less supersymmetric cases.

Let us consider the motion of a rigid string on $S^3_{\gamma}$.
This space can be thought of as a subspace of the
$\gamma$-deformation of $AdS_5\times S^5$ presented above
\begin{equation}
\m_3=0, \quad \phi_3=0 \quad \text{i.e.} \quad \psi=0, \quad
\phi_3=0. \label{3-sphere}
\end{equation}
The relevant part of the $\gamma$-deformed $AdS_5\times S^5$ is
\begin{equation}
ds^2 = -dt^2 + d\theta^2 + G\sin^2\theta d\phi_1^2 +G\cos^2\theta
d\phi_2^2 \label{deformedmetric}
\end{equation}
where $G = \ds\frac{1}{1+\tilde{\gamma}^2
\sin^2\theta\cos^2\theta}$ and due to the series of T-dualities
there is a non-zero component of the B-field
\begin{equation}
B_{\phi_1 \phi_2} = \tilde{\gamma} G \sin^2\theta\cos^2\theta
\end{equation}
We will work in the conformal gauge and thus use the Polyakov action
($T=\frac{\sqrt{\lambda}}{2\pi}$)
\begin{multline}
S = \ds\frac{T}{2}\int d^2\sigma [ -(\partial_{\tau}t)^2 +
(\partial_{\tau}\theta)^2 - (\partial_{\sigma}\theta)^2 +
G\sin^2\theta ((\partial_{\tau}\phi_1)^2 -
(\partial_{\sigma}\phi_1)^2) +G\cos^2\theta
((\partial_{\tau}\phi_2)^2 - (\partial_{\sigma}\phi_2)^2)\\ +
2\gamma G\sin^2\theta\cos^2\theta
(\partial_{\tau}\phi_1\partial_{\sigma}\phi_2-\partial_{\sigma}\phi_1\partial_{\tau}\phi_2)]
\end{multline}
which is supplemented by the Virasoro constraints
\begin{equation}
g_{\mu\nu}\partial_{\tau}X^{\mu}\partial_{\sigma}X^{\nu} = 0
\qquad\qquad\qquad
g_{\mu\nu}(\partial_{\tau}X^{\mu}\partial_{\tau}X^{\nu}+\partial_{\sigma}X^{\mu}
\partial_{\sigma}X^{\nu})
= 0.
\end{equation}
Here $g_{\mu\nu}$ is the metric (\ref{deformedmetric}) and
$X^{\mu}=\{t,\theta,\phi_1,\phi_2\}$. The ansatz
\begin{equation}
t = \kappa\tau \qquad\qquad \theta = \theta(y) \qquad\qquad \phi_1
= \omega_1\tau + \tilde{\phi}_1(y) \qquad\qquad \phi_2  =
\omega_2\tau + \tilde{\phi}_2(y)
\end{equation}
describes the motion of rigid strings on the deformed 3-sphere,
here we have defined a new variable $y=\alpha\sigma+\beta\tau$.
One can substitute the above ansatz in the equations of motion and
use one of the Virasoro constraints to find three first order
differential equations for the unknown functions:
\begin{equation}
\begin{array}{l}
\tilde{\phi_1}' = \ds\frac{1}{\alpha^2 - \beta^2}\left(
\ds\frac{A}{G \sin^2\theta} +\beta\omega_1
-\tilde{\gamma}\alpha\omega_2\cos^2\theta  \right)\\\\
\tilde{\phi_2}' = \ds\frac{1}{\alpha^2 - \beta^2}\left(
\ds\frac{B}{G\cos^2\theta} +\beta\omega_2
+\tilde{\gamma}\alpha\omega_1\sin^2\theta \right)\\\\
(\theta')^2 = \ds\frac{1}{(\alpha^2-\beta^2)^2} [
(\alpha^2+\beta^2)\kappa^2 - \ds\frac{A^2}{G\sin^2\theta} -
\ds\frac{B^2}{G\cos^2\theta}
-\alpha^2\omega_1^2\sin^2\theta-\alpha^2\omega_2^2\cos^2\theta \\\\
+ 2\tilde{\gamma}\alpha
(\omega_2A\cos^2\theta-\omega_1B\sin^2\theta ) ]
\end{array}
\end{equation}
$A$ and $B$ are integration constants and the prime denotes
derivative with respect to $y$. The other Virasoro constraints
provides the following relation between the parameters
\begin{equation}
A \omega_1 + B \omega_2 + \beta\kappa^2=0
\end{equation}
This system has three conserved quantities - the energy and two
angular momenta:
\begin{equation}
\begin{array}{l}
E = 2 T \ds\frac{\kappa}{\alpha} \int_{\theta_0}^{\theta_1}
\ds\frac{d\theta}{\theta'}\\\\
J_1 = 2  \ds\frac{T}{\alpha} \int_{\theta_0}^{\theta_1}
\ds\frac{d\theta}{\theta'} G\sin^2\theta [\omega_1
+\beta\tilde{\phi}_1' +
\tilde{\gamma}\alpha\cos^2\theta\tilde{\phi}_2']\\\\
J_2 = 2  \ds\frac{T}{\alpha} \int_{\theta_0}^{\theta_1}
\ds\frac{d\theta}{\theta'} G\cos^2\theta [\omega_2
+\beta\tilde{\phi}_2' +
\tilde{\gamma}\alpha\sin^2\theta\tilde{\phi}_1']
\end{array}\label{conserved}
\end{equation}
where the integration is performed over the range of the
coordinate $\theta$. In the analysis below we will find solutions
of the above equations and relations between the energy and the
angular momenta for some special values of the parameters. These
solutions include the giant magnon and the single spike solution
on the deformed $S^3$.

The conditions which determine the type of the solution come from the 
requirement of existence of a turning point at $\theta=\pi/2$. This condition sets 
$B=0$ and provides the following choice
\begin{equation}
\begin{array}{c}
(i)\qquad \ds\frac{\kappa^2}{\omega_1^2} = 1 \qquad\qquad
\text{the giant
magnon solution of \cite{Hofman:2006xt}}\\\\
(ii) \qquad\ds\frac{\kappa^2 \beta^2}{\alpha^2 \omega_1^2} = 1
\qquad\qquad \text{the single spike solution of
\cite{Ishizeki:2007we}}
\end{array}
\end{equation}

The dispersion relations in the two cases are as follows.

\paragraph{Giant magnons}\

If we choose $\kappa^2 = \omega_1^2$ (which through the Virasoro
constraint implies $A = -\beta\omega_1$) we get the giant magnon
solution on $S^3_{\gamma}$ found in \cite{Bobev:2006fg,Bobev:2007bm}. The
equations of motion for this case are:
\begin{equation}
\begin{array}{l}
\tilde{\phi}_1' = - \ds\frac{\cos^2\theta}{\alpha^2-\beta^2}
\left(\ds\frac{\beta\omega_1}{\sin^2\theta} +
\tilde{\gamma}\alpha\omega_2+\tilde{\gamma}^2\beta\omega_1 \right)\\\\
\tilde{\phi}_2' = \ds\frac{\beta\omega_2 +
\tilde{\gamma}\alpha\omega_1\sin^2\theta}{\alpha^2 - \beta^2}\\\\
\theta' =
\ds\frac{\alpha\Omega_0}{(\alpha^2-\beta^2)}\ds\frac{\cos\theta}{\sin\theta}
\sqrt{\sin^2\theta -\sin^2\theta_0}
\end{array}\label{eomdefmagnon}
\end{equation}
where we have defined
\begin{equation}
\sin\theta_0 = \ds\frac{\beta\omega_1}{\alpha\Omega_0}
\qquad\qquad \text{and} \qquad\qquad \Omega_0 =
\ds\sqrt{\omega_1^2 - \left(\omega_2 +
\tilde{\gamma}\ds\frac{\beta\omega_1}{\alpha}\right)^2}
\end{equation}
Using the expressions for the energy and the angular momentum
(\ref{conserved}) and equations (\ref{eomdefmagnon}) we find
\begin{equation}
\begin{array}{l}
E - J_1 = 2T\ds\frac{\omega_1}{\Omega_0}\cos\theta_0\\\\
J_2 = 2T\left(\ds\frac{\omega_2}{\Omega_0} +
\tilde{\gamma}\ds\frac{\beta\omega_1}{\alpha\Omega_0}\right)\cos\theta_0
\end{array}
\end{equation}
These expressions lead to the dispersion relation for the giant
magnon solution on $\gamma$-deformed $S^3$ \cite{Chu:2006ae},
\cite{Bobev:2006fg}
\begin{equation}
E-J_1 = \ds\sqrt{J_2^2 + \ds\frac{\lambda}{\pi^2}\cos^2\theta_0}
\end{equation}
In order to make a connection with the spin chain description we
should identify $\cos\theta_0 =
\sin\left(\frac{p}{2}-\pi\beta\right)$, where $p$ is the momentum
of the magnon excitation on the spin chain and $\beta =
\tilde{\gamma}/\sqrt{\lambda}$. So the prediction for the relevant
spin chain dispersion relation is
\begin{equation}
E-J_1 = \ds\sqrt{J_2^2 +
\ds\frac{\lambda}{\pi^2}\sin^2\left(\frac{p}{2}-\pi\beta\right)}
\end{equation}
this relation is invariant under $p\rightarrow p + 2\pi$ and
$\beta\rightarrow \beta+1$ as is required by the spin chain
analysis \cite{Frolov:2005ty}, \cite{Beisert:2005if}.
In \cite{Bykov:2008bj} a detailed analysis of the infinite limit of the charges as well as
finite size corrections is presented. It was argued that in the limit
$\J=J/g\rightarrow\infty$ the dispersion relations does not feel the deformation since
it shows up just as a shift by $\pi\gamma$. It is important however that the deformation produces a non-trivial twist in the boundary conditions for the isometry directions which is proportional to
$\sim \gamma J$. The latter has non-trivial consequences the analysis of which can be seen
in \cite{Bykov:2008bj}.

\paragraph{Single spikes}\

The string profile with one single spike and large winding number is realized when
$\beta^2\kappa^2 = \alpha^2\omega_1^2$ and hence $A = -\frac{\omega_1\alpha^2}{\beta}$
\cite{Ishizeki:2007we}.
It is natural to expect the existence of
rigid string solution solution on $S^3_{\gamma}$ which is the
analogue of the single spike solution on $S^3$ found in
\cite{Ishizeki:2007we}. The equations of motion are
\begin{equation}
\begin{array}{l}
\tilde{\phi}_1' = \ds\frac{1}{\alpha^2-\beta^2}\left(
\beta\omega_1 - \ds\frac{\alpha^2\omega_1}{\beta\sin^2\theta} -
\tilde{\gamma}\alpha\sqrt{\omega_1^2-\Omega_1^2}\cos^2\theta
\right)\\\\
\tilde{\phi}_1' = \ds\frac{1}{\alpha^2-\beta^2}(\beta\omega_2 +
\tilde{\gamma}\alpha\omega_1\sin^2\theta)\\\\
\theta' =
\ds\frac{\alpha\Omega_1}{(\alpha^2-\beta^2)}\ds\frac{\cos\theta}{\sin\theta}
\sqrt{\sin^2\theta
-\sin^2\theta_1}
\end{array}
\end{equation}
where
\begin{equation}
\sin\theta_1 = \ds\frac{\alpha\omega_1}{\beta\Omega_1}
\qquad\qquad\qquad \Omega_1 = \ds\sqrt{\omega_1^2 - \left(\omega_2
+ \tilde{\gamma}\ds\frac{\alpha\omega_1}{\beta}\right)^2}
\end{equation}
The two conserved angular momenta are
\begin{equation}
J_1 = 2 T\ds\frac{\omega_1}{\Omega_1}\cos\theta_1 \qquad\qquad J_2
= -2T\ds\frac{\sqrt{\omega_1^2-\Omega_1^2}}{\Omega_1}\cos\theta_1
\end{equation}
The relation between the conserved charges becomes
\begin{equation}
J_1 = \sqrt{J_2^2 + \ds\frac{\lambda}{\pi^2} \cos^2\theta_1}
\end{equation}
This looks identical to the corresponding expression in the
undeformed case, the dependence on the deformation parameter
$\tilde{\gamma}$ is buried in the definition of $\cos\theta_1$.
In analogy with the giant magnon solution  we can identify
$\cos\theta_1 = \sin\left(\frac{p}{2}-\pi\beta\right)$. 

For the relation between $E$ and $\Delta\phi_1$ we find:
\begin{equation}
E - T\Delta\phi_1 =
\ds\frac{\sqrt{\lambda}}{\pi}\left(\ds\frac{\pi}{2}-\theta_1\right)
-
\tilde{\gamma}\ds\frac{\sqrt{\lambda}}{\pi}\ds\frac{\sqrt{\omega_1^2-
\Omega_1^2}}{\Omega_1}\cos{\theta_1}
\end{equation}
As should be expected in the limit $\tilde{\gamma}\rightarrow 0$
this expression reduces to the one for the single spike solution
on undeformed $S^3$.

\sect{Giant magnon and single spike string solutions on the deformed $T^{1,1}$}

In this section we present the classical solutions in a particular (consistent) subsector 
of the deformed conifold. First we will give a short set up of the beta-deformed conifold\cite{Lunin:2005jy,CatalOzer:2005mr}. Next we consider a solitonic ansatz for giant magnon and single spike classical string solutions and find the explicit form of the solutions. At the end of this Section we briefly comment on the motion of rigid folded strings in the deformed background.


Since the beta-deformed $T^{1,1}$ is known, we will quote here only its final form referring for
instance to \cite{Lunin:2005jy,CatalOzer:2005mr}.
The starting point of the deformation procedure is the metric of $AdS_5\times T^{1,1}$
\eq{
\frac{ds^2}{R^2}=ds^2_{AdS}+\frac{1}{6}\sum\limits_{i=1}^2
(d\theta_i^2+\sin^2\theta_i~d\phi_i^2)+
\frac{1}{9}(d\psi+\cos\theta_1d\phi_1+\cos\theta_2d\phi_2)^2.
}
Here we set the deformation parameter of the squashed sphere to $b=2/3$, i.e. conifold.
Note that there is no B-field. 

According to the procedure described in \cite{Lunin:2005jy,Frolov:2005dj},
the deformed geometry can be obtained by applying T-duality and shift followed by another 
T-duality. The whole procedure can be organized in a single transformation as in \cite{Lunin:2005jy,CatalOzer:2005mr} 
and the result is given by
\ml{
\frac{ds^2}{R^2}=ds^2_{AdS}+G\lb[\frac{1}{6}\sum\limits_{i=1}^2
(G^{-1}d\theta_i^2+\sin^2\theta_i~d\phi_i^2)\right.  \\
\left.+\frac{1}{9}(d\psi+\cos\theta_1d\phi_1+\cos\theta_2d\phi_2)^2+
\tilde\gamma^2\frac{\sin^2\theta_1\sin^2\theta_2}{324}d\psi^2\rb]
}
Due to the T-dualities a non-trivial B-field is generated
\ml{
\frac{B}{R^2}=\tilde\gamma G\lb[(\frac{\sin^2\theta_1\sin^2\theta_2}{36}
+\frac{\cos^2\theta_1\sin^2\theta_2+\cos^2\theta_2\sin^2\theta_1}{54})
d\phi_1\wedge d\phi_2 \right. \\
+\left. \frac{\sin^2\theta_1\cos\theta_2}{54}d\phi_1\wedge d\psi
-\frac{\cos\theta_1\sin^2\theta_2}{54}d\phi_2\wedge d\psi \rb].
}

The conformal factor in the metric and the B-field has the form\footnote{We skip here
the rest of the field content since it will not be used in what follows.}
\eq{
G^{-1}=1+\tilde\gamma^2\lb(\frac{\cos^2\theta_1\sin^2\theta_2+\cos^2\theta_2\sin^2\theta_1}{54}
+\frac{\sin^2\theta_1\sin^2\theta_2}{36}\rb).
\label{def-factor}
}


\subsection{Giant magnon and single spike string solutions}

Let us start with some simplifications of the problem under consideration. The complete solution 
of the non-linear problem is a very complicated task so we will restrict our self to a certain subsector. As in the undeformed case one can  check by direct inspection that the following 
ansatz is a consistent truncation of the complete background.
\eq{
\theta_2=\mathrm{const.} ,\quad \phi_2=\mathrm{const.}
}
To further simplify considerations we choose $\theta_2=0$.
Next we choose the following \textit{ansatz} for solitonic string configurations
\al{
& t=\kappa\tau, \quad \theta_2=0,\quad \phi_2=\mathrm{const.} \notag \\
& \Psi=\o_\psi\tau +\psi(y); \quad \Phi=\o_\phi\tau + \phi(y), \quad \theta=\theta(y),
\label{ansatzII}
}
where $y=c\sigma-d\tau$, $\Psi$ is the $U(1)$ fiber coordinate while $\Phi\equiv\f_1$.

With this choice the metric becomes (we set $R^2=1$)
\eq{
ds^2=- dt^2+\frac{1}{6}d\theta^2+\frac{G}{6}\sin^2\theta d\f^2+
\frac{G}{9}(d\psi+\cos\theta d\f)^2 
\label{metric-trunc}
}
and the B-field takes the form
\eq{
B=G\tilde\gamma\frac{\sin^2\theta}{54}.
\label{b-f}
}
In \eqref{metric-trunc} and \eqref{b-f} the factor $G$ can be read off from \eqref{def-factor}
\eq{
G^{-1}=1+\tilde\gamma^2\frac{\sin^2\theta}{54}.
}
We are looking for solutions with the profile of arc or spike moving along the isometry directions and described by \eqref{ansatzII}. The Lagrangian can be easily deduced from 
(\ref{metric-trunc},\ref{b-f}) and takes the form
\ml{
\L\sim \dot t^2+\frac{1}{6}(-\dot\theta^2+\theta'^2)+
\frac{G}{9}\lb(1+\frac{\sin^2\theta}{2}\rb)(-\dot\Phi^2+\Phi'^2)\\
+\frac{G}{9}(-\dot\Psi^2+\Psi'^2) +\frac{2G}{9}\cos\theta(-\dot\Psi\dot\Phi+\Psi'\Phi')
+2G\tilde\gamma\frac{\sin^2\theta}{54}
\big(\dot\Phi\Psi'-\dot\Psi\Phi'\big).
}
In terms of $\theta,\f$ and $\psi$ it reads off
\ml{
\L\sim \dot t^2+\frac{c^2-d^2}{6}\theta'^2+\frac{G}{9}\Big(1+\frac{\sin^2\theta}{2}\Big)
\Big[-\big(\o_\f-d\f')^2+c^2{\f'}^2\Big]\\
+2\frac{G}{9}\cos\theta\Big[-\big(\o_\psi-d\psi'\big)\big(\o_\f-d\f'\big)+c^2\psi'\f'\Big]
+\frac{G}{9}\Big[-\big(\o_\psi-d\psi'\big)^2+c^2\psi'^2\Big]  \\
+2G\tilde\gamma\frac{\sin^2\theta}{54}c\Big[\big(\o_\f-d\f'\big)\psi'
-\big(\o_\psi-d\psi'\big)\f'\Big].
\label{lagrangianII}
}

It is easy to vary the action and to obtain the equations of motion for $\psi$ and $\f$.
They can be integrated once providing expressions in terms of $\theta$. Explicitly it goes as follows. For $\psi$ we get
\ml{
\p_y\Big\{\frac{2G}{9}\big[d\big(\o_\psi -d\psi'\big)+c^2\psi'\big]
+\frac{2G}{9}\cos\theta\big[d\big(\o_\f -d\f'\big)+c^2\f'\big]\\
+2G\tilde\gamma\frac{\sin^2\theta}{54}c\o_\f\Big\}=0,
}
or
\eq{
\big(c^2-d^2\big)\psi'+d\o_\psi +\cos\theta\big[\big(c^2-d^2\big)\f'+d\o_\f\big]
+\tilde\gamma\frac{\sin^2\theta}{6}c\o_\f=\frac{9A_\psi}{2G}.
\label{eom2-psi1}
}

Analogously, for $\f$ we find
\ml{
\p_y\Big\{\frac{2G}{9}\big(1+\frac{\sin^2\theta}{2}\big)\big[d\big(\o_\f
 -d\f'\big)+c^2\f'\big]
+\frac{2G}{9}\cos\theta\big[d\big(\o_\psi -d\psi'\big)+c^2\psi'\big]\\
-2G\tilde\gamma\frac{\sin^2\theta}{54}c\o_\psi\Big\}=0,
}
or
\eq{
\big(1+\frac{\sin^2\theta}{2}\big)\big[\big(c^2-d^2\big)\f'+d\o_\f\big] +\cos\theta\big[\big(c^2-d^2\big)\psi'+d\o_\psi\big]
-\tilde\gamma\frac{\sin^2\theta}{6}c\o_\psi=\frac{9A_\f}{2G}.
\label{eom2-f1}
}
It is easy to obtain expressions for $\psi'$ and $\f'$ separately, namely
from (\ref{eom2-psi1}) and (\ref{eom2-f1}) we get
\eq{
\big(c^2-d^2\big)\f'+d\o_\f=\frac{3\big(A_\f-A_\psi\cos\theta\big)}{G\sin^2\theta}
+\tilde\gamma\frac{c}{9}\big(\o_\psi+\o_\f\cos\theta\big),
\label{eom2-f2}
}
and
\eq{
(c^2-d^2)\psi'+\o_\psi d=\frac{3A_\psi}{2G}+\frac{3(A_\psi-A_\f\cos\theta)}{G\sin^2\theta}
-\frac{\tilde\gamma c\o_\psi\cos\theta}{9}-\frac{\tilde\gamma c\o_\f}{9}
\big(1+\frac{\sin^2\theta}{2}\big).
\label{eom2-psi2}
}

\paragraph{Virasoro constraints}\

One of the important issues are the Virasoro constraints. In the parameterization we work with
the Virasoro constraints have both, diagonal and off-diagonal components non-trivial.
The diagonal part of the Virasoro constraints consists of $T_{\tau\tau}+T_{\s\s}=0$
\ml{
\frac{1}{6}(\dot\theta^2+\theta'^2)+
\frac{G}{3}\lb(\frac{\sin^2\theta}{2}+\frac{\cos^2\theta}{3}\rb)(\dot\Phi^2+\Phi'^2)\\
+\frac{G}{9}(\dot\Psi^2+\Psi'^2)
+\frac{2G}{9}\cos\theta(\dot\Psi\dot\Phi+\Psi'\Phi')=\kappa^2.
}
or
\ml{
\frac{c^2+d^2}{6}\theta'^2+\frac{G}{9}\Big(1+\frac{\sin^2\theta}{2}\Big)
\Big[\big(\o_\f-d\f')^2+c^2{\f'}^2\Big]
+2\frac{G}{9}\cos\theta\Big[\big(\o_\psi-d\psi'\big)\big(\o_\f-d\f'\big)+c^2\psi'\f'\Big]\\
+\frac{G}{9}\Big[\big(\o_\psi-d\psi'\big)^2
+c^2\psi'^2\Big]=\kappa^2.
\label{virii-diag0}
}
For future use it is convenient to rewrite it in the form
\ml{
\frac{1}{6}\theta'^2+\frac{G}{9}\Big(1+\frac{\sin^2\theta}{2}\Big)
\Big[\f'^2+\frac{\o_\f^2-2d\o_\f \f'}{c^2+d^2}\Big]
+2\frac{G}{9}\cos\theta\Big[\psi'\f'+\frac{\o_\psi\o_\f-d(\o_\f \psi'+\o_\psi\f')}
{c^2+d^2}\Big]\\
+\frac{G}{9}
\Big[\psi'^2+\frac{\o_\psi^2-2d\o_\psi \psi'}{c^2+d^2}\Big]
=\frac{\kappa^2}{c^2+d^2}.
\label{virii-diag1}
}

The off-diagonal part is
\ml{
-\frac{cd}{6}\theta'^2+\frac{G}{9}\big(1+\frac{\sin^2\theta}{2}\big)(\o_\f-d\f')c\f'
+\frac{G}{9}\big(\o_\psi-d\psi')c\psi'\\
+\cos\theta\frac{G}{9}\Big[\big(\o_\psi-d\psi')\f'+(\o_\f-d\f')\psi'\Big]c=0.
\label{virii-off1}
}
This can be rewritten as
\ml{
\frac{1}{6}\theta'^2+\frac{G}{9}\big(1+\frac{\sin^2\theta}{2}\big)(\f'^2-
\frac{\o_\f\f'}{d})
+\frac{G}{9}(\psi'^2-
\frac{\o_\psi\psi'}{d})\\
+\frac{G}{9}\cos\theta\Big[2\psi'\f'-\frac{\o_\psi\f'+\o_\f\psi'}{d}\Big]=0.
\label{virii-off2}
}
Subtracting \eqref{virii-off2} from \eqref{virii-diag1} we find
\ml{
\frac{G}{9}\big(1+\frac{\sin^2\theta}{2}\big)\Big[\o_\f^2
+\frac{(c^2-d^2)\o_\f\f'}{d}\Big]+\frac{G}{9}\Big[\o_\psi^2+
\frac{(c^2-d^2)\o_\psi\psi'}{d}\Big]\\
+\frac{G}{9}\cos\theta\Big[2\o_\psi\o_\f+
\frac{(c^2-d^2)(\o_\psi\f'+\o_\f\psi')}{d}\Big]=\kappa^2.
\label{mixii-vir-sub}
}

Substituting the explicit form of $\f'$ and $\psi'$ from \eqref{eom2-f2} and \eqref{eom2-psi2} 
we find
\eq{
\o_\f A_\f+\o_\psi A_\psi=2\kappa^2 d.
\label{second-vir}
}
This expression puts string restriction on the parameters of the solutions.

\paragraph{Equation of motion for $\theta$}\

The equation of motion for $\theta$ obtained by varying the action is more complicated since it contains the other dynamical variables. It is easier to use another way to obtain it by making use of the Virasoro constraints.
Here we will use the second (off-diagonal) Virasoro constraint to obtain the equation of motion for $\theta$. The latter can be written in the form
\eq{
\theta'^2+
\frac{2G}{3d}\Big\{\big(d\f'-\o_\f\big)\Big[\big(1+\frac{\sin^2\theta}{2}\big)\f'
+\cos\theta\,\psi'\Big]+\big(d\psi'-\o_\psi\big)\Big[\psi'+\cos\theta\,\f'\Big]\Big\}=0.
\label{eom-theta-fin1}
}

From \eqref{eom2-f1} we have
\eq{
\big(c^2-d^2\big)
\Big[\big(1+\frac{\sin^2\theta}{2}\big)\f'
+\cos\theta\,\psi'\Big]=\frac{9A_\f}{2G}+\tilde\gamma c\o_\psi\frac{\sin^2\theta}{6}
-\big(1+\frac{\sin^2\theta}{2}\big)d\o_\f-\cos\theta\,d\o_\psi.
\label{ii-1}
}
From \eqref{eom2-psi1} we find
\eq{
\big(c^2-d^2\big)\Big[\psi'+\cos\theta\,\f'\Big]=
\frac{9A_\psi}{2G}-\tilde\gamma c\o_\f\frac{\sin^2\theta}{6}-d\o_\psi-d\o_\f\cos\theta.
\label{ii-2}
}
On other hand
\eq{
d\f'-\o_\f=\Big\{\frac{3d\big(A_\f-A_\psi\cos\theta\big)}{G\sin^2\theta}
+\tilde\gamma\frac{cd}{9}\big(\o_\psi+\o_\f\cos\theta\big)-c^2\o_\f\Big\}/\big(c^2-d^2\big).
\label{ii-3}
}
and
\ml{
d\psi'-\o_\psi =\Big\{\frac{3dA_\psi}{2G}+\frac{3d(A_\psi-A_\f\cos\theta)}{G\sin^2\theta}
-\frac{\tilde\gamma dc\o_\psi\cos\theta}{9}\\
-\frac{\tilde\gamma dc\o_\f}{9}
\big(1+\frac{\sin^2\theta}{2}\big)-c^2\o_\psi\Big\}/\big(c^2-d^2\big).
\label{ii-4}
}

Substituting into the equation \eqref{eom-theta-fin1} we find
\ml{
\theta'^2+\frac{2G}{3(c^2-d^2)^2}\Big\{
\cdots +c^2\o_\psi (\o_\psi+\o_\f\cos\theta)G^{-1}\\
+c^2\o_\f^2\big(1+\frac{\sin^2\theta}{2}\big)G^{-1}+
c^2\o_\psi\o_\f \cos\theta G^{-1}
\Big\}=0,
\label{eom-theta-fin2a}
}
where $\cdots$ are the terms proportional to $G^{-1}$ and $G^{-2}$ which come from
direct multiplication by $G^{-1}$ in (\ref{ii-1}-\ref{ii-4}). 
The others are organized in $G^{-1}$ as above.

In \eqref{eom-theta-fin2a} the terms in the brackets proportional to $G^{-2}$ are
\eq{
\Big\{\dots_{\vert\sim \frac{1}{G^2}}=\frac{27d}{2G^2\sin^2\theta}\Big[ A_\f\big(A_\f-A_\psi\cos\theta\big)
+A_\psi\big(A_\psi-A_\f\cos\theta\big)+A_\psi^2\frac{\sin^2\theta}{2}
\Big].
\label{1/G2}
}

The terms in the brackets in \eqref{eom-theta-fin2a} proportional to 1/G have contributions from two sources. The contributions coming from $\f$
\ml{
\Big\{\dots_{\f}=\frac{1}{2G\sin^2\theta}\Big[
\sin^2\theta\big[\tilde\gamma cdA_\f\big(\o_\psi+\o_\f\cos\theta\big)
-9c^2\o_\f A_\f \\
-\big(3d^2\o_f-\tilde\gamma dc\o_\psi\big)
\big(A_\f-A_\psi\cos\theta\big)\big]-6d^2\big(A_\f-A_\psi\cos\theta\big)
\big(\o_\f+\o_\psi\cos\theta\big)\Big]
\label{1/Gf}
}
and terms proportional to 1/G coming from $\psi$
\ml{
\Big\{\dots_{\psi}=\frac{-1}{2G\sin^2\theta}\Big[
\sin^2\theta\big[\tilde\gamma cdA_\psi\big(\o_\f+\o_\psi\cos\theta\big)
+\tilde\gamma dc\o_fA_\psi\sin^2\theta +9c^2\o_\psi A_\psi\\
+3d^2A_\psi\big(\o_\psi+\o_\f\cos\theta\big)+\tilde\gamma dc\o_\f
\big(A_\psi-A_\f\cos\theta\big)\big]+6d^2\big(A_\psi-A_\f\cos\theta\big)
\big(\o_\psi+\o_\f\cos\theta\big)\Big]. 
\label{1/Gpsi}
}

To obtain the complete form of the equation we have to add all the terms (\ref{1/G2}-\ref{1/Gpsi})
 and substitute into \eqref{eom-theta-fin2a}.

Let us write down the final form of the equation
\begin{multline}
\theta'^2+\frac{1}{3(c^2-d^2)^2\sin^2\theta}\left\lbrace \left[ \frac{(2c\omega_{\phi}
-{\tilde\gamma} A_{\psi})^2}{4}\right] \sin^4\theta\right. 
+(2c\omega_{\phi}-{\tilde\gamma} A_{\psi})(2c\omega_{\psi}+{\tilde\gamma} A_{\phi})\cos\theta\sin^2\theta\\
+\frac{1}{2}\left[(2c\omega_{\psi}+{\tilde\gamma} A_{\phi})^2+(2c\omega_{\phi}-{\tilde\gamma} A_{\psi})^2+27A_{\psi}^2-\frac{18}{d}(c^2+d^2)(A_{\phi}\omega_{\phi}+A_{\psi}\omega_{\psi})
\right]\sin^2\theta\\
\left.-54A_{\phi}A_{\psi}\cos\theta+27(A_{\psi}^2+A_{\phi}^2)\right\rbrace =0.
\label{3.35}
\end{multline}

For future use we need an expression in terms of $\cos\theta$ only, i.e.
the equation in terms of $\cos\theta$ becomes
\begin{multline}
\theta'^2+\frac{1}{3(c^2-d^2)^2\sin^2\theta}\left\lbrace 
\left[ \frac{(2c\omega_{\phi}-\tilde\gamma A_{\psi})^2}{4}\right] \cos^4\theta\right. -(2c\omega_{\phi}-\tilde\gamma A_{\psi})(2c\omega_{\psi}+\tilde\gamma A_{\phi})\cos^3\theta \\
-\frac{1}{2}\left[(2c\omega_{\psi}+\tilde\gamma A_{\phi})^2+2(2c\omega_{\phi}-\tilde\gamma A_{\psi})^2  +27A_{\psi}^2-\frac{18}{d}(c^2+d^2)(A_{\phi}\omega_{\phi}+A_{\psi}\omega_{\psi})\right]
\cos^2\theta\\
+\left[(2c\omega_{\phi}-\tilde\gamma A_{\psi})(2c\omega_{\psi}+\tilde\gamma A_{\phi})-54A_{\phi}A_{\psi} \right] \cos\theta
+\frac{1}{2}\left[(2c\omega_{\psi}+\tilde\gamma A_{\phi})^2+\frac{3}{2}(2c\omega_{\phi}
-\tilde\gamma A_{\psi})^2\right.\\ \left. \left. +27(3A_{\psi}^2+2A_{\phi}^2)-\frac{18}{d}(c^2+d^2)(A_{\phi}\omega_{\phi}+A_{\psi}
\omega_{\psi})\right]\right\rbrace =0.
\label{3.36}
\end{multline}

\paragraph{The turning point}\

As discussed in the Introduction, we are looking for solutions describing strings with certain
profile, namely, arcs or spikes. Therefore, these must have turning points.
Here we derive the relations following from the condition \eqref{3.35} to have
a turning point at $\theta^\ast=\pi$. 

One can see that the only singular terms at $\theta^\ast=\pi$ are those in the last line
of \eqref{3.35}. To cancel these singularities we must impose some conditions. The last line 
can be written as
\eq{
\frac{27}{\sin^2\theta}\big[
\big(A_\f-A_\psi\cos\theta\big)^2+A^2_\psi\sin^2\theta\big].
\label{turn-1}
}
Then if
\eq{
A_\psi=-A_\f,
\label{turn-2}
}
in the limit $\theta\rightarrow\pi$ the first term in the brackets vanishes as $\sim 0^4$ and the only finite contribution comes from the second term ($=27A_\psi^2$). With \eqref{turn-2} one can safely write down the turning point condition at $\theta^\ast=\pi$
\ml{
\frac{1}{2}(2c\o_\f+\tilde\gamma A_\f)^2 +\frac{1}{2}(2c\o_\psi+\tilde\gamma A_\f)^2
-(2c\o_\f+\tilde\gamma A_\f)(2c\o_\psi+\tilde\gamma A_\f) +\frac{81}{2}A^2_\f \\
=\frac{18}{2d}(c^2+d^2)A_\f(\o_\f-\o_\psi),
\label{turn-3}
}
or
\eq{
81 A^2_\f-18\frac{c^2+d^2}{d}A_\f(\o_\f-\o_\psi)+4c^2(\o_\f-\o_\psi)^2=0.
\label{turn-4}
}
The last equation can be written as
\eq{
\Big(A_\f-\frac{2}{9}d(\o_\f-\o_\psi)\Big)\Big(A_\f-\frac{2c^2}{9d}(\o_\f-\o_\psi)\Big)=0.
\label{turn-5}
}
From \eqref{turn-5} we find two conditions
\eq{
A_\f=\begin{cases}
     \: \frac{2}{9}d(\o_\f-\o_\psi)\quad \text{giant magnon} \\
	\: \frac{2c^2}{9d}(\o_\f-\o_\psi) \quad \text{single spike}
     \end{cases}
\label{turn-6}
}
We remind that the last expression is accompanied by
$$
A_\psi=-A_\f.
$$

\paragraph{The solution}\

Let us introduce for convenience the notations:
\begin{equation}
B_\psi=2c\omega_{\psi}+{\tilde\gamma} A_{\phi},\quad B_\phi=2c\omega_{\phi}+{\tilde\gamma} 
A_{\phi}.
\label{bf-bpsi}
\end{equation} 
In terms of $\cos^2\frac{\theta}{2}=u$ the equation \eqref{3.36}
 can be written in the following form:
\begin{multline}
4u'^2+\frac{1}{3(c^2-d^2)^2}\left\lbrace 4B_\phi^2u^4-8B_\phi(B_\phi+B_\psi)u^3\right. \\
+2\left[B_\phi^2-B_\psi^2+6B_\phi B_\psi-27A_\phi^2+\frac{18}{d}(c^2+d^2)A_{\phi}(\omega_{\phi}-\omega_{\psi}) \right]u^2\\
\left. +2\left[(B_\psi-B_\phi)^2+81A_{\phi}^2-\frac{18}{d}(c^2+d^2)A_{\phi}(\omega_{\phi}
-\omega_{\psi}) \right]u\right\rbrace =0.
\label{eq-final1}
\end{multline}
The turning point condition \eqref{turn-6} in these variables is
\begin{equation}
(B_\psi-B_\phi)^2+81A_{\phi}^2-\frac{18}{d}(c^2+d^2)A_{\phi}(\omega_{\phi}-\omega_{\psi})=0,
\end{equation}
and therefore  the last term in the equation \eqref{eq-final1} vanishes 
\begin{equation}
u'^2=\frac{1}{3(c^2-d^2)^2}\left[ -B_\phi^2u^4+2B_\phi(B_\phi+B_\psi)u^3-(B_\phi^2+2B_\phi B_\psi+27A_\phi^2)u^2\right].
\label{eq-final2}
\end{equation}

A simple analysis analogous to that in \cite{Benvenuti:2008bd} shows that
the equation \eqref{eq-final2} can be written as:
\begin{equation}
u'^2=\frac{B_\phi^2}{3(c^2-d^2)^2}\,u^2\,(\alpha_>-u)(u+\alpha_-),
\end{equation}
where 
\begin{equation}
u'^2=\frac{B_\phi^2}{3(c^2-d^2)^2}\,u^2\,(\alpha_>-u)(u+\alpha_-),
\label{eq-str1}
\end{equation}
and
\begin{align}
&0<\alpha_> = 1+\frac{B_\psi}{B_\phi}\left(1-\sqrt{1-27\frac{A_\phi^2}{B_\psi^2}} \right) <1,
\notag \\
&\alpha_-=|\alpha_<|=-1-\frac{B_\psi}{B_\phi}\left(1+\sqrt{1-27\frac{A_\phi^2}{B_\psi^2}} 
\right)>0,
\label{turning-magnons}
\end{align}
and $0 \leq u \leq \alpha_> <1$.

The solution can be easily obtained and is given by:
\begin{equation}
u(y)=\left( \frac{2\alpha_>\alpha_-}{\alpha_> + \alpha_-}\right)\frac{1}{\cosh\left(|a|\sqrt{\alpha_>\alpha_-}\,\,y \right) -\frac{\alpha_> - \alpha_-}{\alpha_> + \alpha_-}},
\label{solution1}
 \end{equation}
where
\eq{
a^2=\frac{B_\phi^2}{3(c^2-d^2)^2}.
}

Having obtained the solutions with the desired profile one can proceed with the 
dispersion relations.


\sect{Dispersion relations}

In this Section we derive the conserved charges and the corresponding dispersion relations. 
Having obtained the classical string solutions it is easy to compute the conserved charges. 
Due to the specific regime we are working in, namely the very high energies corresponding to 
very long dual operators, some of them are finite but some are divergent. The two cases, giant magnons and single spikes differs in boundary conditions, i.e. the profile of the string 
propagating along the isometry directions.

\subsection{Conserved charges}

Let us start with computing the conserved quantities in the theory.
By definition the conserved momenta corresponding to the isometries are:
\eq{P_\psi = \frac{\partial\L}{\partial(\p_\tau\Psi)},\,\,\,\,P_\phi = \frac{\partial\L}{\partial(\p_\tau\Phi)},\,\,\,\,P_t = \frac{\partial\L}
{\partial(\p_\tau t)}
}

Their explicit form is given by
\al{
& -\frac{2}{T} P_\psi  =-\frac{2G}{9}\Big[\p_\tau\Psi+\cos\theta\p_\tau\Phi +\tilde\gamma
\frac{\sin^2\theta}{6}\p_\sigma\Phi \Big] ,\\
& -\frac{2}{T} P_\phi  =-\frac{2G}{9}\Big[\big(1+\frac{\sin^2\theta}{2}\big)\p_\tau\Phi+\cos\theta\p_\tau\Psi
-\tilde\gamma \frac{\sin^2\theta}{6}\p_\sigma\Psi,\\
& -\frac{1}{T} P_t =\p_\tau t,
}
or, substituting for $\p_{\s,\tau}\Psi$ and $\p_{\s,\tau}\Phi$
\al{
& -\frac{2}{T} P_\psi = \frac{2G}{9}\Big[d\psi'-\o_\psi+\cos\theta(d\f'-\o_\f)-
\tilde\gamma c\frac{\sin^2\theta}{6}\f'\Big], \label{p-psi}\\
& -\frac{2}{T} P_\phi=\frac{2G}{9}\Big[\big(1+\frac{\sin^2\theta}{2}\big)
\big(d\f'-\o_\f\big)+\cos\theta(d\psi'-\o_\psi)+
\tilde\gamma c\frac{\sin^2\theta}{6}\psi'\Big], \label{p-f}\\
& -\frac{1}{T}P_t=\kappa. \label{p-t}
}

The corresponding charges are defined by
\al{
& J_\psi = \int\limits_{-\infty}^\infty \frac{dy}{c} P_\psi =\frac{T}{9}
\int\limits_{-\infty}^\infty \frac{dy}{c}
G\Big[\o_\psi-d\psi'+\cos\theta(\o_\f-d\f')+
\tilde\gamma c\frac{\sin^2\theta}{6}\f'\Big],
\\
& J_\f=\int\limits_{-\infty}^\infty \frac{dy}{c} P_\f =\frac{T}{9}
\int\limits_{-\infty}^\infty \frac{dy}{c}
\Big[\big(1+\frac{\sin^2\theta}{2}\big)
\big(\o_\f-d\f'\big)+\cos\theta(\o_\psi-d\psi')-
\tilde\gamma c\frac{\sin^2\theta}{6}\psi'\Big],
\\
& E =-\int\limits_{-\infty}^\infty \frac{dy}{c} P_t
=T\int\limits_{-\infty}^\infty \frac{dy}{c}\kappa.
}

In the rest of this subsection we will compute explicitly the above charges. Due to the 
specific limit of large quantum numbers some expressions are divergent and we will analyze 
them here. 
In order to obtain the dispersion relations we need to find certain finite combinations out 
of the divergent ones. Below we start this analysis.

\paragraph{Computation of $P_\psi$}\

To obtain the dispersion relations we need the explicit form of the conserved charges.
Let us first find the explicit expression for $P_\psi$
\ml{
-\frac{2}{T}\:
P_\psi=\frac{2G}{9(c^2-d^2)}\Big\{\frac{3dA_\psi}{2G}+\frac{3d(A_\psi-A_\f\cos\theta}
{G\sin^2\theta}
-\tilde\gamma\frac{dc}{9}\big(\o_\f+\o_\psi\cos\theta\big)-\tilde\gamma\frac{dc\o_\f}{9}.
\frac{\sin^2\theta}{2}\\
-c^2\o_\psi
+\frac{3d\cos\theta(A_\f-A_\psi\cos\theta}{G\sin^2\theta}+\tilde\gamma\frac{cd}{9}
\big(\o_\f\cos^2\theta+\o_\psi\cos\theta\big) \\
-c^2\o_\f\cos\theta- \tilde\gamma\frac{\sin^2\theta}{6}c\big(c^2-d^2\big)\f'\Big\}\\
=\frac{2G}{9(c^2-d^2)}\Big\{\frac{9dA_\psi}{2G}-c^2\big(\o_\psi+\o_\f\cos\theta\big)
-\tilde\gamma c\frac{\sin^2\theta}{6}\big[(c^2-d^2)\f'+d\o_\f\big]\Big\}.
\label{p-psi1}
}
The expression in the square brackets we can replace by the expression from \eqref{eom2-f2}. 
The result we find is
\eq{
-\frac{2}{T}\:P_\psi=\frac{2G}{9(c^2-d^2)}\Big\{\frac{9dA_\psi-\tilde\gamma c(
A_\f-A_\psi\cos\theta)}{2G}-c^2\big(\o_\psi+\o_\f\cos\theta\big)
\big[1+\tilde\gamma^2\frac{\sin^2\theta}{54}\big]\Big\}.
\label{p-psi2}
}
One can observe that the expression in the square brackets is exactly $G^{-1}$ so the final 
explicit expression for the momentum $P_\psi$ is
\eq{
-\frac{2}{T}\:P_\psi=\frac{1}{9(c^2-d^2)}\Big[9dA_\psi-\tilde\gamma c
\big(A_\f-A_\psi\cos\theta\big)
-2c^2\big(\o_\psi+\o_\f\cos\theta\big)\Big].
\label{p-psi3}
}

\paragraph{Computation of $P_\f$}\

Here we derive the explicit form of $P_\f$. We start with \eqref{p-f}
\ml{
-\frac{2}{T}\:P_\f=\frac{2G}{9}\Big\{ \big(1+\frac{\sin^2}{2}\big)\big(d\f'-\o_\f\big)
+\cos\theta\big(d\psi'-\o_\psi\big)+ \tilde\gamma c\frac{\sin^2\theta}{6}\psi'\Big\} \\
=\frac{2G}{9}\Big\{\big[\big(1+\frac{\sin^2}{2}\big)\f'+\cos\theta\:\psi'\big]-
\big(1+\frac{\sin^2\theta}{2}\big)\o_\f-\cos\theta\:\o_\psi+
\tilde\gamma c\frac{\sin^2\theta}{2}\:\psi'\Big\}.
\label{p-f1}
}
The expression in the square brackets is given in \eqref{ii-1} and its substitution into
\eqref{p-f1} gives
\ml{
-\frac{2}{T}\:P_\f=\frac{2G}{9(c^2-d^2)}\Big\{\frac{9dA_\f}{2G}+\tilde\gamma
\frac{\sin^2\theta}{6}c
\big[(c^2-d^2)\psi'+d\o_\psi\big]-c^2\Big((1+\frac{\sin^2\theta)}{2}\o_\f \\
 +\cos\theta\o_\psi\Big).
\Big\}
\label{p-f2}
}
The expression in the square brackets is exactly that of \eqref{eom2-psi2}. As result we find
\ml{
-\frac{2}{T}\:P_\f=\frac{2G}{9(c^2-d^2)}\Big\{\frac{9dA_\f+\tilde\gamma c\frac{\sin^2\theta}
{2}A_\psi +\tilde\gamma c(A_\psi-A_\f\cos\theta)}{2G} \\
-c^2\Big(\big(1+\frac{\sin^2\theta}{2}\big)\o_\f+\cos\theta\o_\psi\Big)
\big[1+\tilde\gamma^2\frac{\sin^2\theta}{54}\big]\Big\}.
\label{p-f3}
}
One can recognize in the square brackets the expression for $G^{-1}$ so the final expression 
takes the form
\ml{
-\frac{2}{T}\:P_\f=\frac{1}{9(c^2-d^2)}\Big[9dA_\f-\tilde\gamma cA_\f\cos\theta
+\tilde\gamma c\big(1+\frac{\sin^2\theta}{2}\big)A_\psi \\
 -2c^2\Big(\big(1+\frac{\sin^2\theta}{2}\big)\o_\f+\cos\theta\o_\psi\Big)\Big].
\label{p-f4}
}

As in the undeformed case the momentum $P_\psi$ is linear in $u(y)=\cos^2\theta/2$ while the
momentum $P_\f$ is quadratic. It is expected to have a part analogous to the dispersion 
relations in the undeformed case (with some $\tilde\gamma$ deformations), but we also expect 
to have additional terms. Only the explicit computations can answer the question what is the 
meaning and importance of the deformation.

\paragraph{The angle amplitude}\

From \eqref{eom2-f2} and \eqref{eom2-psi2} it is clear that integrating 
$\psi'$ and $\f'$ we get divergent result. As in the 
undeformed case, one can look for a finite expression combining the two angles $\f'$ and $\psi'$.

To this end we define the following combination
\eq{
\Delta\vf=\int\: dy\frac{\f'-\psi'}{2}.
\label{ang-def1}
}
Now we are going to find explicit expressions for the integrand and analyze 
the eventual divergences.

We start with subtracting \eqref{eom2-f2} and \eqref{eom2-psi2}
\ml{
\big(c^2-d^2\big)\big[\f'-\psi'\big]=
\frac{3(A_\f-A_\psi)(1+\cos\theta)-3A_\psi\frac{\sin^2}{2}}{\sin^2\theta}
\big(1+\frac{\tilde\gamma^2}{54}\sin^2\theta\big)-d(\o_\f-\o_\psi)\\
+\tilde\gamma\frac{c}{9}(\o_\f+\o_\psi) (1+\cos\theta)
+\tilde\gamma\frac{c\o_\f}{9}\frac{\sin^2\theta}{2}.
\label{angle-1}
}
To have uniform description it is better to pass to variable $u=\cos^2\theta/2$ and use
\eq{
1+\cos\theta=2u, \quad \sin^2\theta=4u(1-u).
}
Thus, we find for the ``angle deficit'' the expression
\ml{
\big(c^2-d^2\big)\big[\f'-\psi'\big]=
\frac{6(A_\f-A_\psi)-6A_\psi(1-u)-4d(\o_\f-\o_\psi)(1-u)}{4(1-u)}
+\tilde\gamma\frac{2c}{9}(\o_\f-\o_\psi)u\\
+\frac{\tilde\gamma^2}{9}\big[(A_\f-A_\psi)u-A_\psi(1-u)u\big]
+\tilde\gamma\frac{2c}{9}\o_\f(1-u)u.
\label{angle-2}
}
Using the condition for the turning point \eqref{turn-2} we find
\ml{
\big(c^2-d^2\big)\big[\f'-\psi'\big]=
\frac{18A_\f-4d(\o_\f-\o_\psi)+\big[4d(\o_\f-\o_\psi)-6A_\f\big]u}{4(1-u)}
+\tilde\gamma\frac{2c}{9}(\o_\f-\o_\psi)u\\
+\frac{\tilde\gamma^2}{9}\big[(A_\f-A_\psi)u-A_\psi(1-u)u\big]
+\tilde\gamma\frac{2c}{9}\o_\f(1-u)u.
\label{angle-3}
}
The final expression we will use in what follows is 
\eq{
\big(c^2-d^2\big)\big[\f'-\psi'\big]=\frac{3A_\phi}{1-u}+\frac{3}{2}A_\phi
-d(\omega_\phi-\omega_\psi)+\frac{\tilde\gamma}{9}(2B_\phi+B_\psi)u-\frac{\tilde\gamma}
{9}B_\phi u^2.
\label{ang-final}
}
Let us make a few remarks about the behavior of the angle deficit in the two cases we are 
going to analyze. It is easy to see that integrating $\sim 1/(1-u)$ we will have divergent 
result.
In the magnon case the divergent term $\sim 1/(1-u)$ \textit{vanishes} due to
\eqref{turn-6} and the expression becomes finite. In the single spike case the angle 
deficit is still divergent as it should be (we consider configurations with large winding numbers). In what follows we will consider the two cases separately.

\subsection{Dispersion relations for giant magnons}

In this subsection we will derive the dispersion relations for the giant magnons in the deformed
conifold. The giant magnons are characterized with certain conditions, namely
for giant magnon string solutions we have
$$
 A_\f=\frac{2}{9}d(\o_\f-\o_\psi), \quad A_\psi=-A_\f
$$
which combined with the second Virasoro constraint \eqref{second-vir} gives
\eq{
\kappa=\frac{\o_\f-\o_\psi}{3}
\label{magnon-kappa}
}

Next task is to compute the conserved quantities for this case.

\paragraph{Expression for $P_\psi$}\

The expression for $P_\psi$ in terms of $u$ is
\eq{
-\frac{2}{T}\:P_\psi=-\frac{1}{9(c^2-d^2)}\Big[
9d\Big(A_\f-\frac{2c^2}{9d}\big(\o_\f-\o_\psi\big)\Big)+
\big(4c^2\o_\f+2\tilde\gamma cA_\f\big)\:u\Big].
}
For magnon choice of $A_\f$ \eqref{turn-6} we find 
\eq{
-\frac{2}{T}\:P_\psi=\frac{2}{3}\cdot\frac{\o_\f-\o_\psi}{3}
-\frac{1}{9(c^2-d^2)}\Big(4c^2\o_\f+2\tilde\gamma cA_\f\Big)\:u.
\label{ppsi-magnon1}
}
We note that the first term is exactly $\kappa$, c.f. \eqref{magnon-kappa}.

It is easy now to write the expression for $P_\psi$
\eq{
-\frac{2}{T}\:P_\psi=\frac{2}{3}\cdot\frac{\o_\f-\o_\psi}{3}-
\frac{2c}{9(c^2-d^2)}B_\f\:u.
\label{ppsi-magnon2}
}
It is clear that integrating \eqref{ppsi-magnon2} to obtain the conserved charge we get divergent result. However, the combination
\eq{
E+3J_\psi=\frac{B_\f T}{3(c^2-d^2)}\int dy\: u,
}
is finite since the first term of $P_\psi$ cancels against $\kappa$ from $P_t$.

\paragraph{Expression for $P_\f$}\
 
The next ingredient we need for the dispersion relations is $P_\f$.
We simply substitute the constants for the magnon case in the corresponding expression 
\eqref{p-f4} and obtain
\eq{
-\frac{2}{T}\:P_\f=-\frac{2}{3}\cdot\frac{\o_\f-\o_\psi}{3}
-\frac{2c}{9(c^2-d^2)}\Big[
\Big(2c(\o_\psi+\o_\f)+2\tilde\gamma A_\f\Big)u
-\Big(2c\o_\f+\tilde\gamma A_\f\Big)u^2
\Big].
\label{pf-magnon1}
}
Also, introducing $B_\f$ and $B_\psi$ as defined in \eqref{bf-bpsi} we get
\eq{
-\frac{2}{T}\:P_\f=-\frac{2}{3}\cdot\frac{\o_\f-\o_\psi}{3}
-\frac{2c}{9(c^2-d^2)}\Big[
\big(B_\f+B_\psi\big)u -B_\f\:u^2\Big].
}

One can observe that the finite combination here is
\eq{
E-3J_\f,
}
where the first term cancel $\kappa$ from $\E$.

\paragraph{Expression for $\Delta\vf$}\

As we already mentioned in the last subsection, the angle deficit $\Delta\vf$
defined by \eqref{ang-def1} is finite Let us derive the explicit expression for $\Delta\vf$.

The integrand in \eqref{ang-def1} in the magnon case can be derived by just substituting the
corresponding values for the constants in \eqref{ang-final}
\ml{
\big(c^2-d^2\big)\big[\f'-\psi'\big]
=\frac{2d}{3}(\o_\f-\o_\psi)\frac{u}{1-u}+
\Big(3\tilde\gamma^2 A_\f+2\tilde\gamma c(2\o_\f+\o_\psi)\Big)\frac{u}{9}
-\Big(\tilde\gamma^2 A_\f+2\tilde\gamma c\o_\f\Big)\frac{u^2}{9}\\
=\frac{2}{3}d(\omega_\phi-\omega_\psi)\frac{u}{1-u}+\frac{\tilde\gamma}{9}
(2B_\phi+B_\psi)u- \frac{\tilde\gamma}{9}B_\phi u^2.
}

Since the combinations
\al{
&\frac{E}{T}+3\frac{J_\psi}{T}=\frac{1}{3(c^2-d^2)} B_\phi \int\limits_{-\infty}^\infty u dy\label{com-psi}\\
&\frac{E}{T}-3\frac{J_\phi}{T}=\frac{1}{3(c^2-d^2)} B_\phi \int\limits_{-\infty}^\infty u^2 dy -\frac{1}{3(c^2-d^2)} (B_\phi +B_\psi)\int\limits_{-\infty}^\infty u dy,
\label{com-phi}
}
are finite, one can write the integrand as:
\ml{
\big(c^2-d^2\big)\big[\f'-\psi'\big]=\frac{2}{3}d(\omega_\phi-\omega_\psi)\frac{u}{1-u}+\\+
\tilde\gamma\frac{(c^2-d^2)}{3c}\left[\frac{1}{T}(-P_t)+3\frac{P_\psi}{T} \right] -\tilde\gamma\frac{(c^2-d^2)}{3c}\left[\frac{1}{T}(-P_t)-3\frac{P_\phi}{T} \right].
}

The angle amplitude then takes the following final form
\eq{
\Delta\vf=\int\limits_{-\infty}^\infty dy\frac{\f'-\psi'}{2}
=\frac{d(\omega_\phi-\omega_\psi)}{3(c^2-d^2)}\int\limits_{-\infty}^\infty dy\frac{u}{1-u}+\frac{\tilde\gamma}{6}\left[\frac{E}{T}+3\frac{P_\psi}{T} \right]-\frac{\tilde\gamma}{6}\left[\frac{E}{T}-3\frac{P_\phi}{T} \right].
}

\paragraph{The dispersion relations}\

Let us define the charge densities as
 $E/T=\mathcal E,\quad J_\psi /T=\mathcal J_\psi,\quad J_\phi /T=\mathcal J_\phi$ . 
Then the finite combination of charges take the form:
\begin{align}
&\mathcal E + 3\mathcal J_\psi =\frac{\sqrt{3}}{3} a I_1,\\
&\mathcal E - 3\mathcal J_\phi =\frac{\sqrt{3}}{3} a I_2 - 
\frac{\sqrt{3}}{3} a \frac{(\alpha_> - \alpha_-)}{2} I_1,\\
&\Delta\varphi=\sqrt{(1+\alpha_-)(1-\alpha_>)}\frac{a}{2}I_3+
\frac{\tilde\gamma}{2}(\mathcal J_\psi +\mathcal J_\phi),
\end{align}
where the integrals $I_i$ are given in the Appendix and $\a_-$ and $\a_>$ are defined
in \eqref{turning-magnons}.
Using the explicit form of the integrals $I_i$ (\ref{int-1}-\ref{int-3}) we find:
\begin{align}
&\mathcal E + 3\mathcal J_\psi =\frac{2\sqrt{3}}{3} 
\arccos\left( \frac{\alpha_- -\alpha_>}{\alpha_- +\alpha_>}\right) ,\\
&\mathcal E - 3\mathcal J_\phi =\frac{2\sqrt{3}}{3} \sqrt{\alpha_>  \alpha_-},\\
&\Delta\varphi=\arccos\left(\frac{\alpha_- -\alpha_>}{\alpha_- +\alpha_>}-\frac{2\alpha_> \alpha_-}{\alpha_- +\alpha_>} \right) +\frac{\tilde\gamma}{2}(\mathcal J_\psi +\mathcal J_\phi).\label{delta-phi}
\end{align}
From here we find that the constants $\a_>$ and $\a_-$ are 
related to the charge densities as follows:
\begin{align}
&\sqrt{\alpha_> \alpha_-} =\frac{3}{2\sqrt{3}}(\mathcal E - 3\mathcal J_\phi),\\
&\frac{\alpha_- -\alpha_>}{\alpha_- +\alpha_>}=\cos\left[\frac{3}{2\sqrt{3}}(\mathcal E + 
3\mathcal J_\psi) \right]. 
\end{align}

Simple algebraic calculations\footnote{We use for instance that
$(\a_--\a_>)/(\a_>+\a_-)-\sqrt{\a_>\a_-}\sqrt{1-((\a_--\a_>)/(\a_>+\a_-))^2}=
\cos (\Delta\vf-\tilde\gamma/2(\J_\f+\J_\psi))$.}
lead to
\begin{equation}
\cos\left[\frac{3}{2\sqrt{3}}(\mathcal E + 3\mathcal J_\psi) \right]
-\frac{3}{2\sqrt{3}}(\mathcal E - 3\mathcal J_\phi)
\sin\left[\frac{3}{2\sqrt{3}}(\mathcal E + 3\mathcal J_\psi) \right]
=\cos\big(\Delta\vf-\frac{\tilde\gamma}{2}\big(\J_\f+\J_\psi\big).
\end{equation}

The final form of the dispersion relations in the magnon case is
\begin{equation}
\frac{\sqrt{3}}{2}(\mathcal E - 3\mathcal J_\phi)
=\frac{\cos\left[\frac{\sqrt{3}}{2}(\mathcal E + 3\mathcal J_\psi) \right]
-\cos\big(\Delta\vf-\tilde\gamma/2(\J_\f+J_\psi) \big)}
{\sin\left[\frac{\sqrt{3}}{2}(\mathcal E + 
3\mathcal J_\psi) \right]}.
\label{magnon-disp-rel-fin}
\end{equation}

To close this subsection let us make short comments. First of all, the transcendental 
character of the dispersion relation persists in the deformed background. The deformation 
parameter enters the expression by shifting the angle amplitude by term proportional to 
$\gamma$ times the total spin\footnote{In \cite{Bykov:2008bj} another regime where the $\gamma$ deformation scales to zero is realized}. The BMN and basic giant magnon analysis considered in \cite{Benvenuti:2008bd} can be easily repeated with the same conclusions (up to the $gamma$ shift). Note that each conserved charge depends on the $\gamma$ parameter but this dependence in hidden in the dispersion relations.


\subsection{Dispersion relations for single spike strings}

To obtain the dispersion relation for the single spike strings we have to compute the conserved quantities with the parameters describing strings with large winding number. This requirement
leads to the relations \eqref{turn-6} between the parameters.

Let us start with the condition
for single spike string solutions 
$$
 A_\f=\frac{2c^2}{9d}(\o_\f-\o_\psi),
$$
which combined with the second Virasoro constraint \eqref{second-vir} gives
\eq{
\kappa=\frac{c(\o_\f-\o_\psi)}{3d}.
}
Next, we have to compute all the conserved charges with these adjustments of the parameters.

\paragraph{Expressions for $P_\psi$ and $P_\f$}\

We start with the expression for $P_\psi$. The simple substitution of the above values for the parameters into \eqref{p-psi3} gives
\eq{
-\frac{2}{T} P_\psi = -\frac{1}{9(c^2-d^2)}
\Big[9dA_\f+2c^2(\omega_\psi-\omega_\phi)+2cB_\phi\,u\Big].
\label{p-f-sp1}
}

Analogously, the expression for $P_\f$ derived from \eqref{p-f4} has the form
\eq{
-\frac{2}{T} P_\phi = \frac{1}{9(c^2-d^2)}
\left[9dA_\f+2c^2(\omega_\psi-\omega_\phi)-2c(B_\phi+B_\psi)
\,u+2cB_\phi\,u^2\right].
\label{p-psi-sp1}
}

\paragraph{Expression for $\Delta\vf$}\

The general expression for the angle amplitude
\eq{
\big(c^2-d^2\big)\big[\f'-\psi'\big]=\frac{3A_\phi}{1-u}
+\frac{3}{2}A_\phi-d(\omega_\phi-\omega_\psi)+\frac{\tilde\gamma}{9}
(2B_\phi+B_\psi)u-\frac{\tilde\gamma}{9}B_\phi u^2
}
in the case of a single spike profile the solution takes the form
\eq{
\big(c^2-d^2\big)\big[\f'-\psi'\big]=3A_\phi \frac{u}{1-u}+3\frac{(c^2-d^2)}{c}\kappa+\frac{\tilde\gamma}
{9}(2B_\phi+B_\psi)u-\frac{\tilde\gamma}{9}B_\phi u^2.
\label{angle-def-sp1}
}

\paragraph{The dispersion relations}\

The conserved quantities are
\al{
& \frac{1}{T} P_\psi = \frac{1}{9}\frac{c}{(c^2-d^2)} B_\phi u\\
& \frac{1}{T} P_\phi =\frac{1}{9}\frac{c}{(c^2-d^2)} (B_\phi +B_\psi)u
-\frac{1}{9}\frac{c}{(c^2-d^2)} B_\phi u^2.
}
One can see that the charges obtained by integrating the spins in this case are finite:
\al{
& \frac{1}{T} J_\psi = \frac{1}{9(c^2-d^2)} B_\phi \int\limits_{-\infty}^\infty dy u\\
& \frac{1}{T} J_\phi =\frac{1}{9(c^2-d^2)} (B_\phi +B_\psi) \int\limits_{-\infty}^\infty dy u-\frac{1}{9(c^2-d^2)} B_\phi \int\limits_{-\infty}^\infty dy u^2.
}
Therefore, the total momentum density
\eq{
\frac{1}{T} P_\psi+\frac{1}{T} P_\phi=\frac{1}{9}\frac{c}{(c^2-d^2)} (2B_\phi +B_\psi)u-\frac{1}{9}\frac{c}{(c^2-d^2)} B_\phi u^2,\label{sum}
}
also defines a finite charge.

As in the undeformed case, the situation with the angle amplitude is more tricky.
From \eqref{angle-def-sp1} and (\ref{p-f-sp1},\ref{p-psi-sp1} one finds
\eq{
\big(c^2-d^2\big)\big[\f'-\psi'\big]
=3A_\phi \frac{u}{1-u}+3\frac{(c^2-d^2)}{c}\left(-\frac{P_t}{T} \right) +\tilde\gamma \frac{(c^2-d^2)}{c}\left(\frac{1}{T} P_\psi+\frac{1}{T} P_\phi \right). 
\label{delta-sp1}
}
The angle amplitude then is given by
\eq{
\Delta\vf=\int\limits_{-\infty}^\infty dy\frac{\f'-\psi'}{2}=\frac{3A_\phi}{2(c^2-d^2)} \int\limits_{-\infty}^\infty dy\frac{u}{1-u}+\frac{3}{2}\left(\frac{E}{T} \right) +\frac{\tilde\gamma}{2}\left(\frac{J_\psi}{T} +\frac{J_\phi}{T} \right).
\label{delta-sp2}
}
We showed explicitly that the angle amplitude itself is divergent (as is the energy), but
there exists a finite combination that can be used to find dispersion relations.
From \eqref{delta-sp2} we see that the following combination is a finite:
\eq{
\Delta\delta\equiv\Delta\vf-\frac{3}{2}\left(\frac{E}{T} \right)-\frac{\tilde\gamma}{2}\left(\frac{J_\psi}{T} +\frac{J_\phi}{T} \right)=\frac{3A_\phi}{2(c^2-d^2)} \int\limits_{-\infty}^\infty dy\frac{u}{1-u}.
\label{delta-sp3}
}

It is useful to introduce again the charge densities
$E/T=\mathcal E,\,\,\,\, J_\psi /T=\mathcal J_\psi,\,\,\,\, J_\phi /T=\mathcal J_\phi$, 
which are found to be
\begin{align}
&\mathcal J_\psi =\frac{\sqrt{3}}{9} a I_1,\\
&\mathcal J_\phi =-\frac{\sqrt{3}}{9} a I_2 
 + \frac{\sqrt{3}}{9} a \frac{(\alpha_> - \alpha_-)}{2} I_1,\\
&\Delta\delta\equiv\Delta\varphi- \frac{3}{2}\mathcal E -\frac{\tilde\gamma}{2}
(\mathcal J_\psi +\mathcal J_\phi)=\sqrt{(1+\alpha_-)(1-\alpha_>)}\frac{a}{2}I_3.
\end{align}
The substitution of the explicit form of the integrals $I_i$ gives
\begin{align}
&\mathcal J_\psi =\frac{2\sqrt{3}}{9} \arccos\left( \frac{\alpha_- -\alpha_>}{\alpha_- +\alpha_>}\right) ,\\
&\mathcal J_\phi =-\frac{2\sqrt{3}}{9} \sqrt{\alpha_>  \alpha_-},\\
&\Delta\delta=\arccos\left(\frac{\alpha_- -\alpha_>}{\alpha_- +\alpha_>}-\frac{2\alpha_> \alpha_-}{\alpha_- +\alpha_>} \right) .\label{delta-delta}
\end{align}

One can again express the constants $\a_>$ and $\a_-$ in terms of the charge densities
\begin{align}
&-\sqrt{\alpha_> \alpha_-} =\frac{3\sqrt{3}}{2}\mathcal J_\phi,\\
&\frac{\alpha_- -\alpha_>}{\alpha_- +\alpha_>}
=\cos\left(\frac{3\sqrt{3}}{2}\mathcal J_\psi \right).
\end{align}
Using that
\begin{equation}
\frac{\alpha_- -\alpha_>}{\alpha_- +\alpha_>}-\sqrt{\alpha_> \alpha_-} 
\sqrt{1-\left(\frac{\alpha_- -\alpha_>}{\alpha_- +\alpha_>}\right) ^2}=\cos\Delta\delta.
\end{equation}
we find
\begin{equation}
\cos\left(\frac{3\sqrt{3}}{2}\mathcal J_\psi \right)+\frac{3\sqrt{3}}{2}
\mathcal J_\phi \sin\left(\frac{3\sqrt{3}}{2} \mathcal J_\psi \right)=\cos\Delta\delta.
\end{equation}

The final form of the dispersion relations is
\begin{equation}
-\frac{3\sqrt{3}}{2}\mathcal J_\phi =\frac{\cos\left(\frac{3\sqrt{3}}{2}
\mathcal J_\psi \right)-\cos\Delta\delta}{\sin\left(\frac{3\sqrt{3}}{2} 
\mathcal J_\psi \right)}.
\label{disp-sp}
\end{equation}

The transcendental character of the dispersion relations persists as expected. The non-trivial shift of the angle amplitude is of the same form as in the case of giant magnons and therefore has an universal form.

\sect{Conclusions}

In this paper we have studied the problem of existence certain class solitonic solutions of strings in the beta-deformed $T^{1,1}$ which is the base of the conifold. The latter is an important example of a string dual of gauge theory with less than $\N=4$ supersymmetry and has many interesting applications.

To set up the notations and make the paper more self contained,
first we give a short review of the magnon and single spike solutions in the undeformed case as well as the magnon and spiky solutions in $\gamma$-deformed sphere $S^3_\gamma$. In the next Section we present our original results, which can be summarized as follows. In Section 3 we derive and analyze
the classical string solitons of the magnon and single spike strings type for a subsector of the $\gamma$-deformed conifold. In the next section we obtain the dispersion relations for the classical solution found in the previous section. The results show that the dispersion relations are of the same transcendental type as the ones in the undeformed case \cite{Benvenuti:2008bd}. The explicit dependence on $\gamma$ shows up as a shift of the (generalized) angular amplitude - a behavior familiar from the studies of most supersymmetric case of strings in deformed $AdS_5\times S^5$ background \cite{Bobev:2005cz,Bobev:2006fg,Chu:2006ae,Bobev:2007bm}.

There are two essential differences from the known result of giant magnon and single spike strings in $AdS_5\times S^5$. First one is that the dispersion relations related the conserved charges in a transcendental way in both cases - undeformed and $\gamma$-deformed  conifold. This indicates that if there are integrable structures on the conifold they will be much more complicated than the known from the most supersymmetric case. The second difference is that although the dispersion relations explicitly ``feels'' the $\gamma$-deformation through a shift in the angular extend of the magnon profile or winding number, in our case it has qualitatively new feature. 
While in the case of sphere the shift was just by $\gamma\pi$\footnote{Note that actually we also used the identification $\tilde\gamma=\sqrt{\a}\gamma$.} and
one can turn on the regime of large charges in which classically the $\gamma$ term scales to zero
\cite{Bykov:2008bj}, here this shift is quite different. As shown in \eqref{magnon-disp-rel-fin} and \eqref{disp-sp}, it is proportional to the total momentum, which in our regime of validity is of the order of energy, i.e. very large. It seems that in our case this contribution cannot be made vanishing and they have exactly the form of the non-trivial twist of the boundary conditions. In any case it is interesting that the shift involves also the conserved charges, a feature which deserves to be thoroughly analyzed.

An important issue to pursue is the search for integrable structures. This is an important open question which, if positive, would have 
important contribution to the understanding of the string/gauge theory 
duality. Another direction is to extend the considerations to the whole $T^{1,1}$ which is less ambitious but also important.

An important issue to pursue is the search for integrable 
structures. This is an important open question which, if , would have 
important contribution to the understanding of the string/gauge theory 
duality. Another direction is to extend the analysis in this paper to  
dynamics on the full  $T^{1,1}$ manifolds\cite{Dimov:}.

Finally, it is know that there exist an one-parameter family of $AdS_5\times X^5$ 
solutions which interpolates between the Klebanov-Witten background and the
Pilch-Warner background \cite{Pilch:2000ej}. 
It would be interesting to look for solitonic solutions in 
Pilch-Warner background more over that not too much is known about quasiclassical 
strings in this background \cite{Dimov:2003bh}.

\section*{Acknowledgments}
We thank N.~Bobev for critically reading the draft and for various comments.
This work was supported in part by the Austrian Research Fund FWF \# P19051-N16
NSFB VU-F-201/06 and DO 02-257.


\begin{appendices}
\renewcommand{\theequation}{\thesection.\arabic{equation}}

\sect{Useful formulae}

\paragraph{Some integrals}\

Let us write down some useful integrals used in the calculations. 
Before that we stress on the follolwing point.
In accordance with the equations of motion we are dealing with solitary wave solution 
$u(y)=\cos^2\theta/2$.
The turning point $\theta(y=0)=\theta_0$ corresponds to $u(y=0)=\alpha_> =\cos^2\frac{\theta_0}{2}$ while the turning point $\theta(y=\infty)=\pi$ corresponds to $u(y=\infty)=0$. Note that 
when $y$ increases from $0$ to $\infty$, $\theta(y)$ increase from $\theta_0$ to $\pi$ ($0<\theta_0\leqslant \theta(y) \leqslant\pi$), and hence the function $u(y)=\cos^2\frac{\theta(y)}{2}$ decrease from $\alpha_>$ to $0$ ($1> \alpha\geqslant u(y)\geqslant 0$).

In computing cinserved charges we need three integrals.
First integral entering the calculations of the dispersion relations is
\eq{
I_1=\int\limits_{-\infty}^\infty u\:dy.
}

To calculate the integral we use \eqref{eq-str1} and obtain
\eq{
I_1=\frac{2}{|a|}\arctan\left(\frac{2\,\sqrt{\alpha_> \alpha_-}}{\alpha_- - \alpha_>}\right)\\
=\frac{2}{|a|}\arccos\left( \frac{\alpha_- -\alpha_>}{\alpha_- +\alpha_>}\right)\\
=\frac{4}{|a|}\arccos\sqrt{\frac{\a_-}{\a_>+\a_-}}
\label{int-1}
}

Another integral appearing in our calculations is
\begin{multline}
I_2 = \int\limits_{-\infty}^\infty\,u^2 dy
=\frac{2}{|a|}\frac{(\alpha_> - \alpha_-)}{2}\arctan
\left(\frac{2\,\sqrt{\alpha_> \alpha_-}}{\alpha_- - \alpha_>}
\right) +\frac{2}{|a|}\sqrt{\alpha_> \alpha_-}=\\
=\frac{2}{|a|}\frac{(\alpha_> - \alpha_-)}{2}
\arccos\left( \frac{\alpha_- -\alpha_>}{\alpha_- +\alpha_>}
\right)+\frac{2}{|a|}\sqrt{\alpha_> \alpha_-} 
\label{int-2}
\end{multline}

We also used
\begin{multline}
I_3 = \int\limits_{-\infty}^\infty\,\frac{u}{1-u}dy\\
=\frac{2}{|a|}\frac{1}{\sqrt{(1+\alpha_-)(1-\alpha_>)}}
\arctan\left(\frac{2\,\sqrt{\alpha_> \alpha_- 
(1+\alpha_-)(1-\alpha_>)}}{\alpha_- - \alpha_> - 2\alpha_> \alpha_-}\right) =\\
=\frac{2}{|a|}\frac{1}{\sqrt{(1+\alpha_-)(1-\alpha_>)}}
\arccos\left(\frac{\alpha_- -\alpha_>}{\alpha_- +\alpha_>}
-\frac{2\alpha_> \alpha_-}{\alpha_- +\alpha_>} \right) .
\label{int-3}
\end{multline}

\paragraph{Some relations}\

There is a relation between $I_1$ and $I_2$, which has the form
\eq{
I_2=\frac{\a_>-\a_-}{2}I_1-\frac{1}{a}\sqrt{\a_>\a_-}.
\label{int-rel1}
}

Also, we have the relations
\begin{equation}
a=\frac{B_\phi}{\sqrt{3}(c^2-d^2)},\,\,\,\,
\frac{(\alpha_> - \alpha_-)}{2}=1+\frac{B_\psi}{B_\phi},\,\,\,\,
\sqrt{(1+\alpha_-)(1-\alpha_>)}=\frac{3\sqrt{3}A_\phi}{B_\phi}.
\end{equation}


\end{appendices}


\end{document}